\documentclass[prb,aps,twocolumn,notitlepage]{revtex4-1}
\usepackage{amssymb}
\usepackage{CJK}
\usepackage{textcomp}
\usepackage{pdfsync}
\usepackage{amsmath}
\usepackage{amsfonts}
\usepackage{graphicx}
\usepackage{bm}
\usepackage{epstopdf}
\usepackage{subfigure}
\usepackage{comment}
\usepackage{verbatim}
\usepackage{my_symbols}
\usepackage{color}
\usepackage{upgreek}
\usepackage{tikz}
\usetikzlibrary{arrows}

\newcommand{\bmu}{\ensuremath{\boldsymbol{\mu}}}

\begin{document}

\begin{CJK*}{UTF8}{gbsn} 

\title{Current-induced spin wave excitation in Pt\textbar YIG bilayer}
\author{Yan Zhou$^{1,2}$}
\email[Corresponding author:~]{yanzhou@hku.hk}
\author{H. J. Jiao$^{3}$, Y. T. Chen$^{3}$}
\author{Gerrit E. W. Bauer$^{3,4}$}
\author{Jiang Xiao (è§æ±)$^{5,6}$}
\email[Corresponding author:~]{xiaojiang@fudan.edu.cn}
\affiliation{$^1$Department of Physics, The University of Hong Kong, Hong Kong, China \\
$^2$Center of Theoretical and Computational Physics, Univ. of Hong Kong, Hong Kong, China\\
$^3$Kavli Institute of NanoScience, Delft University of Technology, Delft, The Netherlands\\
$^4$Institute for Materials Research and WPI-AIMR, Tohoku University, Sendai, Japan \\
$^5$Department of Physics and State Key Laboratory of Surface Physics, Fudan University, Shanghai, China \\
$^6$Center for Spintronic Devices and Applications, Fudan University, Shanghai, China
}

\date{\today}

\begin{abstract}
We develop a self-consistent theory for current-induced spin wave excitations in normal metal\textbar magnetic insulator bilayer structures. We compute the spin wave dispersion and dissipation, including dipolar and exchange interactions in the magnet, the spin diffusion in the normal metal, as well as the surface anisotropy, spin-transfer torque, and spin pumping at the interface. We find that: 1) the spin transfer torque and spin pumping affect the surface modes more than the bulk modes; 2) spin pumping inhibits high frequency spin-wave modes, thereby red-shifting the excitation spectrum; 3) easy-axis surface anisotropy induces a new type of surface spin wave, which reduces the excitation threshold current and greatly enhances the excitation power. We propose that the magnetic insulator surface can be engineered to create spin wave circuits utilizing surface spin waves as information carrier.
\end{abstract}
\maketitle
\end{CJK*}

\section{Introduction}
The rapid development of nanoscale science and technology has opened the way for the new interdisciplinary research field known as magnonics. Magnonic devices utilize propagating spin waves instead of particle currents to transmit and process information in periodically patterned magnetic nanostructures, such as domain walls, magnetic vortices and antivortices, magnetic nanocontacts \etc Magnonic devices potentially combine the advantages of fast speed, easy and wideband tunability, and compactness with compatibility with complementary metal-oxide-semiconductor process. \cite{Kruglyak2010}

A complete magnonic circuit consists of a spin wave injector, a spin wave detector, and a functional medium through which the spin waves propagate and may be manipulated. Due to their exceptionally low magnetic damping, electrically insulating ferro or ferrimagnets  are believed to be  suitable for spin wave transmission line. \cite{kajiwara_detection_2011,khitun_non-volatile_2011} Spin waves can propagate much larger distances in magnetic insulator compared to both spin wave and particle-based spin currents in ferromagnetic metals. A recent experiment has shown that spin Hall spin currents in a normal metal can effectively excite a wide range of spin wave modes by the spin transfer torque in magnetic insulator that is in contact with a normal metal with strong spin-orbit coupling. \cite{kajiwara_transmission_2010} The spin wave detection is made possible through the spin pumping and inverse spin Hall effect.\cite{saitoh_conversion_2006} The magnetic insulator functions as the spin wave transmission medium, inside which different modes of spin waves can propagate. In addition to the conventional bulk/volume modes, a new type of surface spin wave mode due to easy-axis surface anisotropy (EASA) have been recently predicted \cite{xiao_spin-wave_2012} and confirmed.\cite{da_silva_enhancement_2013} The EASA surface waves differ in nature from the magnetostatic surface waves (MSW) mode described by the Damon-Eshbach theory. Because EASA surface waves are strongly localized at the surface, they are  strongly susceptible to the effects of spin transfer torques (STT) and spin pumping (SP), but only weakly absorb microwaves. Da Silva \etal indeed observed such behavior in a recent experiment.\cite{da_silva_enhancement_2013}

In our early study of spin wave excitation in the Pt\textbar YIG system,\cite{xiao_spin-wave_2012,XYB_spin-wave_2013} we were mainly concerned with the magnetization dynamics, disregarding the details of spin transport in the normal metal and SP. SP  affects surface modes more strongly than bulk modes. In a recent theoretical study, it was shown that SP enhances the damping of YIG surface modes  more than that of the bulk modes. \cite{Kapelrud_2013} Due to  spin-transfer torque and spin pumping, the spin transport in the metal and the magnetization dynamics are coupled. So far, all studies have been focusing on one side of the story assuming the other side to be granted. The spin current in the metal has been assumed to be fixed in order to study the magnetization dynamics in magnetic insulators. \cite{xiao_spin-wave_2012,XYB_spin-wave_2013,Kapelrud_2013} The spin transport  in the metal  was studied in detail for a static magnetization of the insulator. \cite{Chen2013} In this paper, we present a complete theory in which  the spin transport and magnetization dynamics are treated on equal footing.  

This paper is organized as follows. In Section \ref{sec:theory}, we present the full theory of current-induced spin wave excitation in Pt\textbar YIG system. Section \ref{sec:SW_analytical} and \ref{sec:results} are devoted to the analytical and numerical results for the spin wave dispersion and dissipation, as well as their dependence on various material parameters including surface anisotropy, spin transfer torque and spin pumping \etc  We conclude in Section \ref{sec:discussions} with a summary of the major results and reflect on the potential technological applications. 

\section{Theory}
\label{sec:theory}


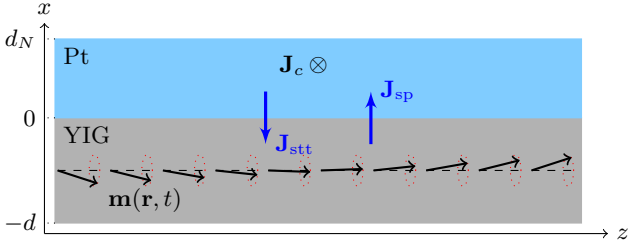
\begin{figure}[t]
\centering
\begin{tikzpicture}[scale=0.7]
        \definecolor{Ptcolor}{rgb}{0.5, 0.8, 1}

        \filldraw[gray!60] (0,-1) rectangle (10,1);
        \node[below right] at (0,1) {YIG};
        \foreach \a in {0,...,9} {
        \draw[dotted,red] (\a+0.75,0) ellipse (.1 and .3);
        \draw[thick,->] (\a+0.05,0) -- +(-18+4*\a:0.8);
        \draw[thin,dashed] (\a+0.05,0) -- +(0:0.8); }
        \node at (-18:1.8) {$\mm(\rr,t)$};

        \filldraw[Ptcolor] (0,1) rectangle (10,2.5);
        \node at (5,2)  {$\otimes$};        
        \node at (4.5,2)  {$\JJ_c$};        
        \node[below right] at (0,2.5) {Pt};
        \draw[very thick,blue,->,>=latex'] (4,1.5) -- (4,0.5) node[right,blue]{$\JJ_{\rm stt}$};
        \draw[very thick,blue,->,>=latex'] (6,0.5) -- (6,1.5) node[right,blue]{$\JJ_{\rm sp}$};


        \draw[->] (-0.2,-1.2) -- (10.5,-1.2) node[right] {$z$};
        \draw[->] (-0.2,-1.2) -- (-0.2,2.8) node[above] {$x$};

        \draw[thin,dotted,-] (0,1) -- (-0.2,1) node[left] {$0$};
        \draw[thin,dotted,-] (0,-1) -- (-0.2,-1) node[left] {$-d$};
        \draw[thin,dotted,-] (0,2.5) -- (-0.2,2.5) node[left] {$d_N$};   \end{tikzpicture}
\caption{(Color online) An electrically insulating magnetic film of thickness $d$ with magnetization $\mm(\rr,t)$ ($\|\hzz$ at equilibrium) in contact with a normal metal of thickness $d_N$, with translational symmetry in the  \textit{y-z} plane. A spin current $\JJ_{\rm s}$ polarized along $\hzz$ is generated in the normal metal due to the spin Hall effect from the applied charge current $J_{\rm c}$ and absorbed by the ferromagnet. $\JJ_{\rm sp}$ is the SP current due to the magnetization dynamics at the interface.}
\label{fig:setup}
\end{figure}

In this section, we present our theory for the spin transport and spin wave excitation in a normal metal (N) - ferromagnetic insulator (FI) bilayer structure as shown in \Figure{fig:setup}, in which the FI is in-plane magnetized with the equilibrium magnetization along the $\hzz$-direction.  

\subsection{Spin transport in normal metal}
\label{sec:Pt_theory}

We assume an electric field $\EE = E_{y}\hyy$ applied in N along $\hyy$. $\JJ_c = \sigma \EE = J_c\hyy$ the charge current, with $\sigma$ the electric conductivity of N. Due to the spin Hall effect, a spin current polarized along $\hzz$ flows in $-\hxx$ direction: $\JJ_{\rm sH} = \theta_H J_c\hzz$ with $\theta_H$ the spin Hall angle of N. This spin Hall current induces a spin accumulation $\bmu(x)$ in N, which satisfies the spin-diffusion equation
\begin{equation}
\label{eqn:sde}
\nabla^{2}\bmu(x)=\frac{\bmu(x)}{\lambda^{2}},
\end{equation}
where $\lambda$ is the spin-flip length in N. The spin current inside N is the sum of the spin diffusion current and the spin Hall current
\begin{equation}
\label{eqn:Jsx}
\JJ_{s}(x)=-\frac{\sigma}{2e}\frac{\partial \bmu(x)}{\partial x}-\theta_HJ_c\hzz.  
\end{equation} 
Spin-conserving boundary conditions require that $\JJ_{s}(x)$ is continuous at the interfaces $x=0$ and $x=d_N$. Thus,
\begin{equation}
\label{eqn:Jbc}
\JJ_{s}(d_N)=0, \quad \JJ_{s}(0) = \JJ_{s0}.
\end{equation}
$\JJ_{s0}$ is the spin current flowing through the N\textbar FI interface, which includes  the STT current $\JJ_{\rm stt}$ generated by the spin accumulation in N on the magnetization in FI and the SP current $\JJ_{\rm sp}$ from FI to N:
\begin{align}
\label{eq_js_if}
&\JJ_{s0} = \JJ_{\rm stt}+\JJ_{\rm sp} \nn
&= {e\ov h}g_r\bigb{\mm(0)\times \midb{\mm(0)\times \bmu(0)} -\hbar \mm(0)\times \dmm(0) },
\end{align}
with $g_{r}$ the real part of the mixing conductance per area for the N\textbar FI interface. In \Eq{eq_js_if}, $\mm$ and $\bmu$ take the value at the interface ($x = 0$). The imaginary part of the mixing conductance is disregarded in the following. 

The solution for $\bmu(x)$ satisfying the spin diffusion equation \Eq{eqn:sde} and boundary condition \Eq{eqn:Jbc} is given by 
\begin{equation}
\bmu(x) = {2e\lambda\ov\sigma}
{\smlb{\JJ_{\rm sH}+\JJ_{s0}}\cosh{d_N-x\ov\lambda}-\JJ_{\rm sH}\cosh{x\ov\lambda}
\ov  \sinh{d_N\ov\lambda}}.
\end{equation}
By plugging the above expression  into the second equation of \Eq{eqn:Jbc}, we find the interfacial value of $\bmu(0)$ and thus $\JJ_{s0}$:
\begin{equation}
\JJ_{s0} 
= {e\ov h}g'_r\midb{ \mm(0)\times(\mm(0)\times\bmu_s^0) - \hbar\mm(0)\times\dmm(0)},
\end{equation}
where $\bmu_s^0 = (2e\lambda/\sigma)\theta_HJ_c\tanh(d_N/2\lambda)\hzz$ is the spin accumulation at the interface due to the spin Hall current alone, and
\begin{equation}
g'_r 
= {g_r\ov 1+{2\lambda e^2\ov h\sigma}g_r\coth{d_N\ov\lambda}}
\end{equation}
is the renormalized mixing conductance taking into account the effect of diffusive spin current back-flow in N. \cite{Tserkovnyak2005}

The interfacial spin current $\JJ_{s0}$ exerts the STT and SP torques on $\mm$:
\begin{subequations}
\label{eqn:torque}
\begin{align}
    \btau_{\rm stt} &= g'_r {e\lambda\theta_H J_c\ov 2\pi\sigma} 
    \tanh{d_N\ov 2\lambda} 
    \mm\times\smlb{\mm\times\hzz}{\delta}(x) \nn
    &\equiv\ \tau_{\rm stt}\mm\times\smlb{\mm\times\hzz} {\delta}(x),\\
    \btau_{\rm sp} &=  -{\hbar\ov 4\pi}g'_r
        { \mm\times \dmm}~{\delta}(x) 
    \equiv -{\tau_{\rm sp}\ov\omega_0}\mm\times \dmm ~{\delta}(x).
\end{align}
\end{subequations} 

\begin{figure}[t]
        \includegraphics[height=0.44\columnwidth]{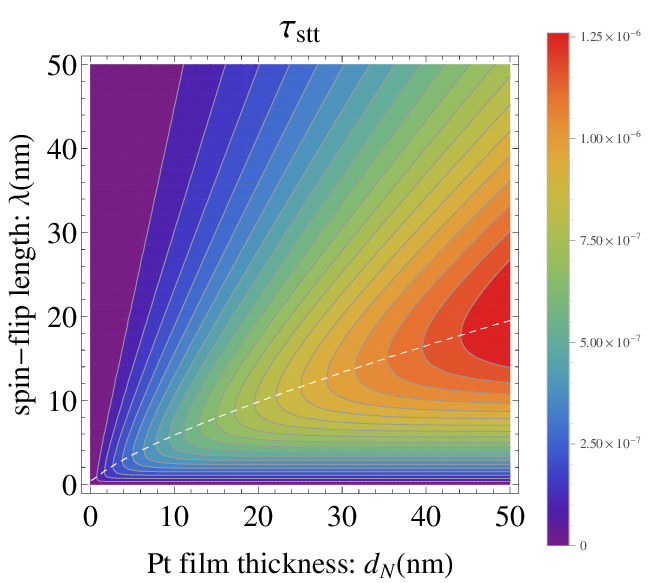}  
        \includegraphics[height=0.44\columnwidth]{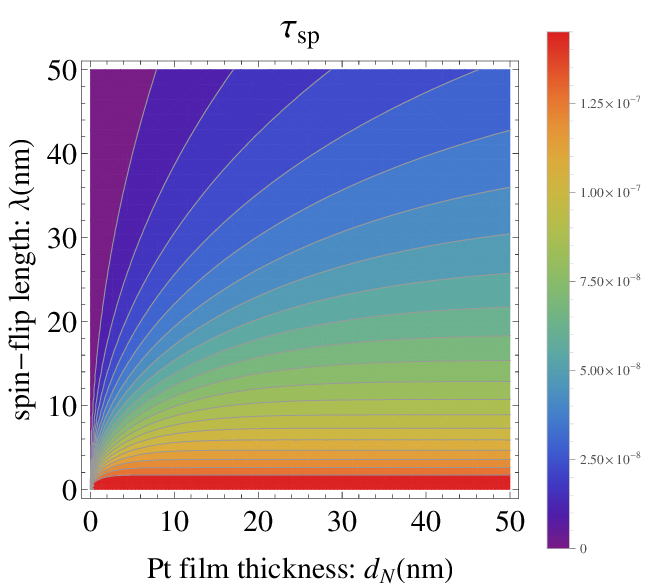}                
    \caption{(Color online) The contour plot of $\tau_{\rm stt}$ (at $J_c = 10^{11}$A/m$^2$, left) and $\tau_{\rm sp}$ (right) in \Eq{eqn:torque} vs. film thickness $d_N$ and spin diffusion length $\lambda$ for parameters given in Table \ref{tab:param} and $g_r = 10^{18}$/m$^2$. The dashed curve on the left panel shows the maximum of $\tau_{\rm stt}$ for fixed film thickness $d_N$. }
  \label{fig:torque}
\end{figure}

\Figure{fig:torque} shows the dependence of the pre-factors of these two torques on the film thickness $d_N$ and spin diffusion length $\lambda$. In the left panel of \Figure{fig:torque}, we see that for a fixed film thickness $d_N$, the STT depends non-monotonically on $\lambda$ and has a maximum value for an intermediate value (indicated by the dashed line). The reason for this is the following: when $\lambda\ra 0$, the spin Hall current cannot build up any spin accumulation, thus there can be no STT; when, on the other hand, $\lambda\ra\infty$, \Eq{eqn:sde} is solved by $\bmu(x)=ax+b$, which means $\JJ_s(x) = \mbox{const}$. However, at the top surface $\JJ_s(d_N) = 0$, therefore the spin current has to vanish everywhere. Both $\JJ_{\rm stt}$ and $\JJ_{\rm sp}$ vanishes, because the above argument is valid for both $\dmm = 0$ and $\dmm \neq 0$. For the SP, the right panel of \Figure{fig:torque}, the behavior is easy to understand. For $\lambda\ra 0$, the SP is maximal because N becomes an ideal spin sink. As $\lambda\ra\infty$, there is no spin flip mechanism in N, so the pumped spin current accumulates in N and causes a back flow spin current, which cancels the pumped spin current.

\subsection{Spin wave excitation in magnetic insulators}
\label{sec:sw_model}

The spatially dependent dynamics of the magnetization unit vector $\mm(\rr,t)$  is described by the Landau-Lifshitz-Gilbert-Slonczewski (LLGS) equation\cite{sun_spin-current_2000,xiao_macrospin_2005,Zhou2013}: 
\begin{equation}
    \dmm = -{\gamma}~\mm{\times}\HH_{\rm eff} + {\alpha}~\mm{\times}\dmm  + 
    {{\gamma}\ov M_s}\smlb{\btau_{\rm stt}+\btau_{\rm sp}},
\label{eqn:llgs}
\end{equation}
where the effective field $\HH_{\rm eff}=\HH_0 + \HH_s + {A_{\rm ex}\ov {\gamma}}{\nabla}^2\mm + \hh$ includes the external magnetic field $\HH_0$, the surface anisotropy field $\HH_s = {2K_1\ov M_s}(\mm{\cdot}\nb)\nb$, the exchange field $\HH_{\rm ex} = {A_{\rm ex}\ov {\gamma}}{\nabla}^2\mm$, and the dipolar magnetic field $h$ due to $\mm(\rr,t)$. Here $\nb$ is the outward normal as seen from the ferromagnet which can be the easy or hard axis, depending on the sign of the anisotropy constant $K_1$. $A_{\rm ex}$ and ${\alpha}$ are the exchange and  Gilbert damping constants, respectively. 

We include the SP in our model thereby extending our earlier studies of spin-wave excitation in magnetic insulators by the STT.\cite{xiao_spin-wave_2012} The spin-conservation boundary conditions for $\mm$ at $x = 0$ and $-d$:\cite{gurevich_magnetization_1996} 
\begin{subequations}
\label{eqn:mbc}
\begin{align}
\mbox{at } x = 0: \quad&
\mm{\times}{{\partial}\mm\ov{\partial}\nb} - k_s(\mm{\cdot}\nb)\mm{\times}\nb \\
&+ k_j\mm{\times}(\mm\times\hzz) + {k_p\ov\omega_0}\hzz{\times}\dmm = 0, \nn
\mbox{at } x = -d: \quad&
\mm{\times}{{\partial}\mm\ov{\partial}\nb} = 0 ,
\end{align}
\end{subequations}
with ${\partial}\mm/{\partial}\nb~{\equiv}~(\nb{\cdot}{\nabla})\mm$ and $K_s = {\int}_{0^-}^{0^+} K_1 dx$. We convert surface anisotropy, spin current, and SP parameters into effective wave numbers by defining: 
\begin{equation}
\label{eqn:wavevector}
k_s = {2{\gamma}K_s\ov A_{\rm ex}M_s},\quad
k_j = {\frac{{\gamma\tau_{\mathrm{stt}}}}{A_{\rm ex}M_s}},\quad
k_p = {\frac{{\gamma\tau_{\mathrm{sp}}}}{A_{\rm ex}M_s}}.
\end{equation}

\begin{table}[t]
        \centering
        \footnotesize \begin{tabular}{c|l|l|c|l|l} \hline
        Param.                  & YIG                           & Unit          & 
        Param.                  & Pt                            & Unit          \\ \hline  
        $M_s$                   & $^a1.56{\times}10^5$          & A/m           &    
        $\sigma$                & $^e1.16{\times}10^{6}$        & A/Vm          \\          
        $\alpha$                & $^a6.7{\times}10^{-5}$        & -             & 
        $\lambda$ &  $^e2$      & nm                                            \\  
        $g_r$                   & $^b10^{16}~{\sim}~10^{19}$    &1/m$^2$        &
        $\theta_H$&0.08         & -                                             \\      
        $K_s$                   & $^c10^{-4}$                   & J/m$^2$       & 
        &                       &                                               \\  
        $A_{\rm ex}$            & $^d8.97{\times}10^{-6}$       & m$^2$/s       & 
        &                       &                                               \\  
        $\gamma$                & $1.76{\times}10^{11}$         & 1/(Ts)       & 
        &                       &                                               \\  
        $\omega_0 = \gamma H_0$ & $^d17.25$                     & GHz           & 
        &                       &                                               \\  
        $\omega_M=\gamma \mu_0M_s$      & $^d34.5$              & GHz           & 
        &                       &                                               \\ \hline
        $d$                     & $0.61$                        & $\upmu$m      & 
        $d_N$                   & $10$                          & nm            \\ \hline
        \end{tabular}
        \caption{Parameters for YIG.
        $^a$Ref. \onlinecite{kajiwara_transmission_2010},
        $^b$Ref. \onlinecite{kajiwara_transmission_2010,Burrowes2012,czeschka_scaling_2011},             
        $^cK_s$ = $0.01 {\sim} 0.1$ erg/cm$^2$ or $10^{-5} {\sim} 10^{-4}$ J/m$^2$, 
        Ref. \onlinecite{yen_magnetic-surface_1979, ramer_effects_1976},  
        $^d$Ref. \onlinecite{de_wames_dipole-exchange_1970}, 
        $^e$Ref. \onlinecite{Chen2013}, 
        $^f$Ref. \onlinecite{kajiwara_transmission_2010}.
        }
        \label{tab:param}
\end{table}


Compared to our previous work,\cite{xiao_spin-wave_2012} we now establish the relation between spin wave vector $k_j$ and the experimentally controlled parameter, \ie the charge current density. For example, the bulk excitation threshold  $k_c = {\alpha}({\omega}_0+{\omega}_M/2)d/A_{\rm ex}$ corresponds to a charge current of $6.6\times10^{11}$A/m$^2$ at $g_r=5.9\times10^{17} /\upmu$m.   

The bulk magnetization inside the film ($-d<x<0$) satisfies the LLG equation: 
\begin{equation}
    \dmm = -{\gamma}~\mm{\times}\midb{\HH_0 + {A_{\rm ex}\ov {\gamma}}{\nabla}^2\mm + \hh} + {\alpha}~\mm{\times}\dmm,
\label{eqn:llg}
\end{equation}
where the dipolar magnetic field $\hh(\rr,t)$ obeys Maxwell's equations in the quasi-static approximation: 
\begin{subequations}
\label{eqn:h}
\begin{align}
\mbox{everywhere: }
0 &= {\nabla}{\times}\hh(\rr), \label{eqn:h1} \\
-d \le x \le 0: 
0 &= {\nabla}{\cdot}\midb{\hh(\rr)+{\mu}_0M_s\mm(\rr)}, \label{eqn:h2} \\
x < -d \qor x > 0: 
0 &= {\nabla}{\cdot}\hh(\rr), \label{eqn:h3}
\end{align}
\end{subequations}
with boundary conditions
\begin{subequations}
\label{eqn:hbc}
\begin{align}
    \hh_{y,z}(0^-) = \hh_{y,z}(0^+), &\quad 
    \bb_x(0^-) = \bb_x(0^+), \\
    \hh_{y,z}(-d^-) = \hh_{y,z}(-d^+), &\quad
    \bb_x(-d^-) = \bb_x(-d^+). 
\end{align}
\end{subequations}
\Eqss{eqn:mbc}{eqn:hbc} completely describe what is called dipolar-exchange spin waves. The method described above extends  De Wames and Wolfram's \cite{de_wames_dipole-exchange_1970} and Hillebrands' \cite{hillebrands_spin-wave_1990} by including the current-induced STT and SP. 

Because of the translational symmetry in the lateral direction, we may assume that the scalar potential is the plane wave: 
\begin{equation}
    {\psi}(x,y,z,t) = {\sum}_{j=1}^3\midb{a_je^{iq_x^{(j)}x}+b_je^{-iq_x^{(j)}(x+d)}} e^{-i\qq{\cdot}\sS} e^{i{\omega}t}
\label{eqn:psi}
\end{equation}
where $\sS = (y, z)$ is the in-plane position and $\qq = (q_y, q_z) = q(\sin{\theta},\cos{\theta})$ with $q = \abs{\qq}$ an in-plane wave vector and $\theta$ the angle between the wave vector $\qq$ and the magnetization equilibrium $\hzz$. $a_j, b_j$ are six coefficients to be determined by the six boundary conditions in \Eqs{eqn:mbc}{eqn:hbc}, which can be transformed into a set of linear equations:  
\begin{equation}
    M(\qq, {\omega}) \smatrix{a_j \\ b_j} = 0,
\label{eqn:Mab}
\end{equation}
where $M(\qq, {\omega})$ is a $6{\times}6$ matrix depending on the material parameters and injected spin current: ${\omega}_0, {\alpha}, k_s, k_j$. The dipolar-exchange spin wave dispersion is determined by the condition that the determinant of the coefficient matrix vanishes: $\abs{M(\qq, {\omega})} = 0 \Ra {\omega}(\qq)$. The corresponding solution of \Eq{eqn:Mab} for $a_j, b_j$ gives the spin wave amplitude profile according to \Eq{eqn:psi}, from which we also see that the spin wave is amplified when 
\begin{equation}
\imm{{\omega}(\qq)} < 0,
\label{eqn:exc}
\end{equation}
which is used as criterium for spin wave excitation with wave vector $\qq$.

\section{Analytical results}
\label{sec:SW_analytical}

The inclusion of the dipolar fields complicates the problem significantly. Nevertheless, it is still possible to obtain approximate analytical expressions of the complex dispersion relation $\omega(\qq)$ for the dipolar-exchange spin waves for the few special cases: 1) the bulk modes for $\theta=\pi/2$; 2) the magnetostatic surface wave  for $\theta=\pi/2$; 3) the surface spin wave mode induced by easy-axis surface anisotropy (EASA) at zero wave-length limit of $q = 0$. While the real part has been  studied quite well before, the imaginary part characterizing the dispersion and excitation of spin waves is usually disregarded and focus of the present study. All analytical expressions in this section are obtained by expanding the relevant matrix $M(\qq,\omega)$ to leading order in: $\alpha$, $k_s$, $k_j$, and $k_p$.

\subsection{Bulk modes for $\theta = \pi/2$}

Assuming weak surface anisotropy ($A_{\rm ex}k_s^2 {\ll} 2\omega_0+\omega_M$) and long wave length limits, the complex eigen-frequency for the $n$th bulk mode reads 
\begin{align}
    \label{eqn:wn} 
    &{\omega}_n = \sqrt{ {\omega}_{nq} ({\omega}_{nq} + {\omega}_M) } \\
    &- {A_{\rm ex}k_s\ov d}
    {{\omega}_{nq}\ov \re{{\omega}_n}} \smlb{ 1 - {{\omega}_{nq}+{\omega}_M\ov 2{\omega}_{nq}+{\omega}_M}
    \sqrt{A_{\rm ex}k_s^2\ov {\omega}_{nq}+{\omega}_0+{\omega}_M} }^{-1}    \nn
   &+i\midb{\left({\alpha}+2{A_{ex}k_p\ov{\omega}_0d}\right)\smlb{{\omega}_{nq} + {{\omega}_M\ov 2}} 
   -\alpha A_{\rm ex}{k_s\ov d} + 2A_{\rm ex}{k_j\ov d}} \nonumber
\end{align}
with ${\omega}_{nq} = {\omega}_0 + A_{\rm ex}[q^2 + (n{\pi}/d)^2]$ and $n=1,2,\dots$.  $\re{\omega_n}$, the real part of the eigenfrequency, decreases with increasing surface anisotropy $k_s$. $\im{\omega_n}$ gives the information about the dissipation (or damping), which includes the contributions from Gilbert damping ($\alpha$ terms), spin current injection ($k_j$ term), and SP ($k_p$ term). For example, the $2A_{\rm ex}k_p^2/\omega_0d$ is the enhanced damping due to SP effect and the $2A_{\rm ex}k_j/d$ is the effect of STT. As expected, both terms are inversely proportional to the film thickness $d$ because both STT and SP are interfacial effect. The spin wave excitation condition $\im{\omega_n} < 0$ leads to  the threshold current for exciting the bulk modes for $\theta = \pi/2$.  

\subsection{Magnetostatic surface wave for $\theta = \pi/2$}

Magnetostatic surface wave (MSW) is a dipolar spin wave mode that exists for $qd \lesssim 1$ at $\theta = \pm\pi/2$. The complex eigen frequency for MSW at ${\theta} = {\pi}/2$ is 
\begin{align}
    {\omega}_{\rm MSW} &= \sqrt{\smlb{{\omega}_0 + {{\omega}_M\ov 2}}^2- {{\omega}_M^2\ov 4}e^{-2qd}}\nn
    &+ i\midb{\left({\alpha}+{A_{ex}k_p\ov{\omega}_0d}\right)\smlb{{\omega}_0+{{\omega}_M\ov 2}} + A_{\rm ex}{k_j\ov d}}.
\label{eqn:wMSW}
\end{align}
Comparing \Eq{eqn:wMSW} for the MSW and \Eq{eqn:wn} for bulk modes, the effect of STT and SP on the former is half of that on bulk modes. It is because the MSW magnetization for $qd\lesssim 1$ has almost constant amplitude over the thickness (\ie a surface wave with  long decay length, see the thick purple curve in \Figure{fig:SP_disper_ks0}(b) below), while the magnetization for bulk modes oscillates as a cosine function (see the thin curves in \Figure{fig:SP_disper_ks0}(b) below). The total magnetization of MSW ($\propto d$ for $qd\lesssim 1$) is therefore twice as large as the total magnetization of the bulk modes ($\propto d/2$ because of the average of a cosine function is 1/2), which reduces the effect of the STT and SP by one half. As before, the threshold current for exciting the magnetostatic surface wave can be derived using the spin wave excitation condition $\im{\omega_{\rm MSW}} < 0$ for $\theta = \pi/2$.

\subsection{EASA induced surface spin wave mode at $q = 0$}

In Ref. \onlinecite{xiao_spin-wave_2012}, the EASA was found to induce a new type of surface spin wave mode, whose penetration depth $d_s$ is inversely proportional to the strength of EASA: $d_s \propto 1/k_s$. In order to understand this EASA surface wave better, we study the limit $d\ra {\infty}$, \ie the magnetic film is semi-infinite and $b_j = 0$ in \Eq{eqn:psi}. Focusing for simplicity on vanishing in-plane wave-vector $\qq = (q^y, q^z) = 0$, the scalar potential can be written as: 
\begin{equation}
    {\psi}(\rr) = {\sum}_{j=1}^2a_je^{iq_jx}e^{i{\omega}t}
    \label{eqn:psis}
\end{equation}
where
\begin{equation}
q_j({\omega}) = 
-i\sqrt{{\omega}_0+\half {\omega}_M {\pm} \sqrt{{\omega}^2+\quarter{\omega}_M^2} {\pm} i{\alpha}{\omega}
\over A_{\rm ex}}
\end{equation}
are negatively imaginary with $\abs{q_1}{\gg} \abs{q_2}$. Imposing the boundary conditions from \Eq{eqn:mbc} at $x = 0$,  $\abs{M(\qq, {\omega})} = 0$ leads to (up to the first order in $k_j$):
\begin{widetext}
\begin{equation}
        0 = 2q_1q_2(q_1+q_2) +ik_s\midb{(q_1+q_2)^2+{{\omega}_M\ov A_{\rm ex}}} 
+ 4k_j{\omega\ov A_{\rm ex}}-2 k_p {\omega\ov \omega_0 }(q_1+q_2)(q_1+q_2+ik_s), 
\label{eqn:det}
\end{equation}
whose solution is the complex eigenfrequencies ${\omega}_S$ for the EASA surface wave. By expanding \Eq{eqn:det} up to the leading orders in $\alpha, k_j, k_p$, and assuming $A_{\rm ex}k_s^2 \ll 2\omega_0+\omega_M$, we have:
\begin{align}
&{\omega}_{\rm S} = \sqrt{{\omega}_0({\omega}_0+{\omega}_M)} \nn
&+ i\smlb{\omega_0+{\omega_M\ov 2}}
\midb{\alpha 
+ {4A_{\rm ex}k_sk_j\omega_0\ov (2{\omega}_0+{\omega}_M)^2}
\smlb{1+k_s\sqrt{A_{\rm ex}(\omega_0+2\omega_M)^2\ov(2\omega_0+\omega_M)^3}} 
+ {2A_{\rm ex}k_sk_p\ov 2\omega_0+\omega_M}
\smlb{1+k_s{\sqrt{A_{\rm ex}(\omega_0+\omega_M)^2\ov (2\omega_0+\omega_M)^3}}}
}.
\label{eqn:wS}
\end{align}
$\im{\omega_{\rm S}} < 0$ leads to:
\begin{align}
        J_{\rm th} = 
        -{\sigma\coth{d_N\ov 2\lambda}\ov 2\theta_H\lambda e}
        \midb{
        { \alpha \pi A_{\rm ex}M_s\ov g'_r\gamma}
        \smlb{ {(2{\omega}_0+{\omega}_M)^2\ov k_sA_{\rm ex}{\omega}_0} 
        - {{\omega}_0+2{\omega}_M\ov {\omega}_0}\sqrt{2{\omega}_0+{\omega}_M\ov A_{\rm ex}} 
}
        +\hbar
        \smlb{ \omega_0+{\omega_M\ov 2} - {k_s\ov 2}\sqrt{A_{\rm ex}\omega_M^2\ov 2\omega_0+\omega_M}}
        }
        .
\label{eqn:jc}
\end{align}
\end{widetext}
The first term of \Eq{eqn:jc} gives the threshold current that compensates the Gilbert damping $\alpha$ for the EASA surface wave of penetration depth $d_s\propto 1/k_s$ (from the first term in the first square bracket). The second term of \Eq{eqn:jc} compensates the SP enhanced damping. 

\begin{figure}[b]
        \includegraphics[height=0.5\columnwidth]{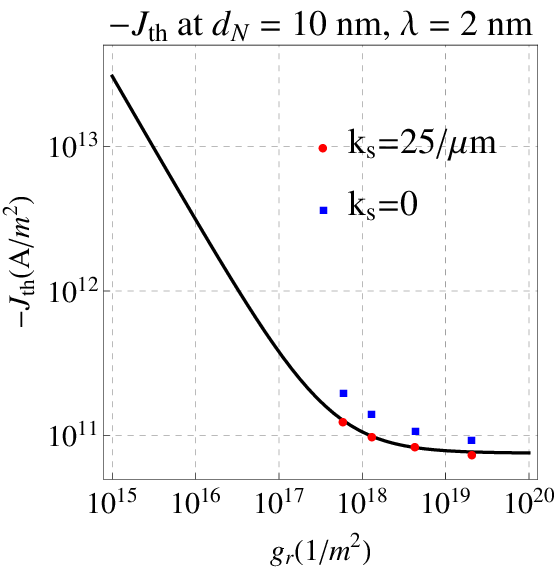}\hspace{3mm}                
        \includegraphics[height=0.5\columnwidth]{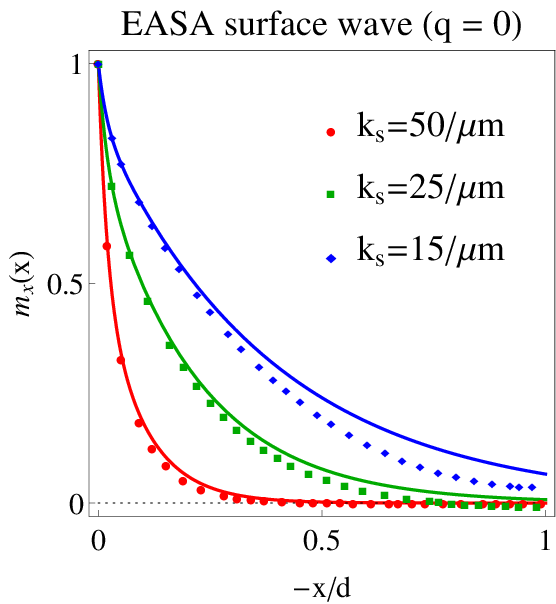}
    \caption{(Color online) 
Left: $J_{\rm th}$ in \Eq{eqn:jc} vs. the mixing conductance $g_r$ (log-log scale) for $k_s = 25/\upmu$m with $d_N = 10$ nm and $\lambda = 2$ nm. The dots are the threshold current obtained from numerical calculations below for $k_s = 25/\upmu$m and $k_s = 0$. 
Right: The magnetization profiles for the EASA surface wave for various $k_s$ values. The solid curves are plotted using \Eq{eqn:mEASA} for a semi-infinite film. The dots are obtained by numerical calculations for $d = 0.61\upmu$m.    
}
  \label{fig:Jc}
\end{figure}

Since $J_{\rm th}$ in \Eq{eqn:jc} is the threshold current for EASA surface wave at $q = 0$, so it actually provides a upper bound for the overall threshold current for the spin wave excitation. However, the excitation threshold current for the EASA surface wave is well below that of other spin wave modes in many cases (\ie for not too small $k_s$), $J_{\rm th}$ in \Eq{eqn:jc} is the overall threshold current for  spin wave excitation in a Pt\textbar YIG bilayer. \Figure{fig:Jc} shows this threshold current as a function of mixing conductance $g_r$. When $g_r$ is not too large (such that $g'_r \simeq g_r$), the threshold current approximately decreases linearly with $g_r$: $J_{\rm th}\propto 1/g_r$, because the STT approximately increases linearly with $g_r$ (see the linear part of left panel in \Figure{fig:Jc}). However, when $g_r$ is large, $g'_r \simeq 1$, then $J_{\rm th}$ is independent of $g_r$, and $J_{\rm th}$ reaches its lower bound (see the flat part of the left panel in \Figure{fig:Jc}). Overall, we expect $J_{\rm th}$ given by \Eq{eqn:jc} to work well as the overall threshold current for intermediate $k_s$. It does not work for small $k_s$, because the penetration depth of EASA surface wave is too long, and the other modes actually have lower threshold current. For larger $k_s$, \Eq{eqn:jc} simply does not work because it is derived assuming small $k_s$.

We may also calculate the spin wave profile for the EASA surface wave. Using \Eq{eqn:wS}
\begin{subequations}
\begin{align}
    q_1&=-i\sqrt{2{\omega}_0+{\omega}_M\ov A_{\rm ex}} \\
    q_2&=-i{\omega_0 k_s\ov 2\omega_0+\omega_M}\smlb{ 1 + k_s\sqrt{A_{\rm ex}({\omega}_0+{\omega}_M)^2\ov (2{\omega}_0+{\omega}_M)^3}}. 
\end{align}
\end{subequations}
Since $q_{1,2}$ are both negative imaginary, the corresponding spin waves in \Eq{eqn:psis} are localized near the surface. The spin wave profile (the $x$ component) for the EASA surface wave for a semi-infinite film is approximately given by:
\begin{equation}
m_x(x) = {(q_1+ik_s)e^{iq_2x}-(q_2+ik_s)e^{iq_1x}\ov q_1-q_2}.
\label{eqn:mEASA}
\end{equation}
Since $\abs{q_1}\gg\abs{q_2}$, the penetration depth is mostly determined by $q_2$: $d_s \propto 1/iq_2 \propto 1/k_s$ for small $k_s$. The spin wave profile in \Eq{eqn:mEASA} is compared with the numerical calculation in the left panel of \Figure{fig:Jc}. The agreement is quite good except for locations near the bottom surface ($x/d\ra 1$) because \Eq{eqn:mEASA} is calculated for semi-infinite films, while the numerical data are computed for a thin film of finite thickness $d = 0.61\upmu$m. The deviation at $x/d\ra 1$ reflects the bottom surface (at $x = -d$) influence on the EASA surface wave localized at the top surface at $x = 0$. Not surprisingly, the effect of the bottom surface is more obvious for the EASA surface wave that is less confined (smaller $k_s$).

 
\section{Numerical results}
\label{sec:results}

\begin{figure}[t]
        \includegraphics[width=\columnwidth,trim= 60 35 70 20, clip=true]{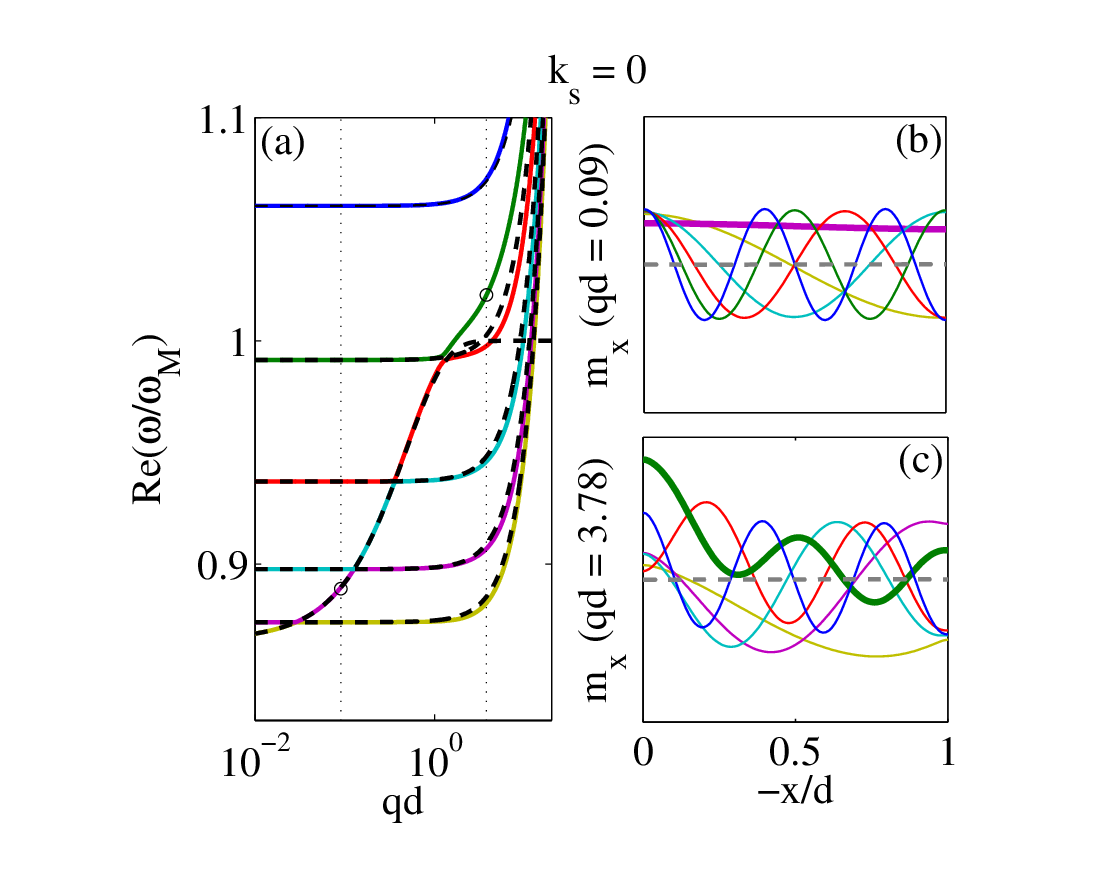}        
    \caption{(Color online) Spin wave dispersion (left) and profiles (right) in the absence of surface anisotropy ($k_s = 0$) at $\theta = \pi/2$ (or $\qq\perp\mm$). Left: spin wave dispersion (a), the solid lines (different colors denote different bands) are calculated from the numerical solution of \Eq{eqn:Mab}, and the dashed lines are plotted using the analytical expressions given by the real parts of \Eqs{eqn:wn}{eqn:wMSW}. Right: spin wave profiles ($m_x$ component) at $qd = 0.09$ (b) and $qd = 3.78$ (c). The colors in (b, c) match that in (a). The thick purple/green mode in (b)/(c) is for the point enclosed with circle in (a) on the purple/green band. }
        \label{fig:SP_disper_ks0}
\end{figure}

\begin{figure}[t]
        \includegraphics[width=\columnwidth,trim= 60 35 70 20, clip=true]{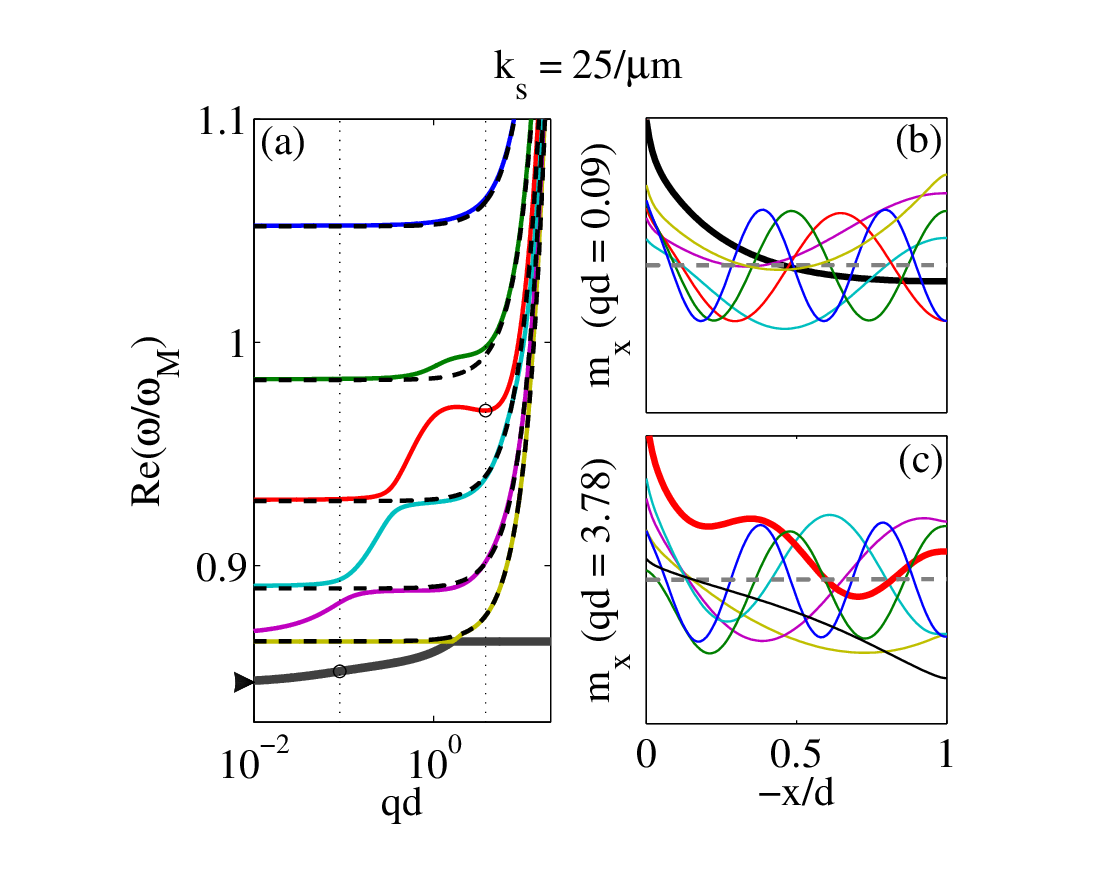}        
    \caption{(Color online) Same as \Figure{fig:SP_disper_ks0} but with easy-axis surface anisotropy ($k_s = 25/\upmu$m). Left: spin wave dispersion (a), the solid lines are calculated from the numerical solution of \Eq{eqn:Mab}, while the dashed lines and the $\blacktriangleright$ symbol are plotted using the analytical expressions given by the real parts of \Eqs{eqn:wn}{eqn:wS}. A new (black) band appears due to the easy-axis surface anisotropy. Right: spin wave profiles ($m_x$ component) at $qd = 0.09$ (b)  and $qd = 3.78$ (c). The colors in (b, c) match that in (a). The thick black/red mode in (b)/(c) are for the points enclosed by a circle in (a) on the black/red band. }
        \label{fig:SP_disper_ks25}
\end{figure}

In this Section, we discuss the effects of the STT and SP on the spin wave excitation. Because of their interfacial character, both STT and SP are more effective for surface spin wave modes. In the absence of STT, the surface spin wave modes have larger damping compared to the bulk modes. When an STT is applied, the surface spin wave modes are easier to excite as well. 


We show the numerical results on the spin wave dispersion as well as the spin wave profiles with different types of surface anisotropy, followed by the corresponding spin wave dissipation affected by the STT and SP. The spin wave excitation power spectrum discussed at the end shows a dramatic effect of EASA and the associated surface wave. If not stated otherwise, the numerical results in this section are calculated for an in-plane magnetized YIG thin film capped with Pt as pictured in \Figure{fig:setup} with geometry and material parameters given in Table \ref{tab:param}.

\begin{figure*}[t]
        \includegraphics[width=1\textwidth,trim= 110 90 110 0, clip=true]{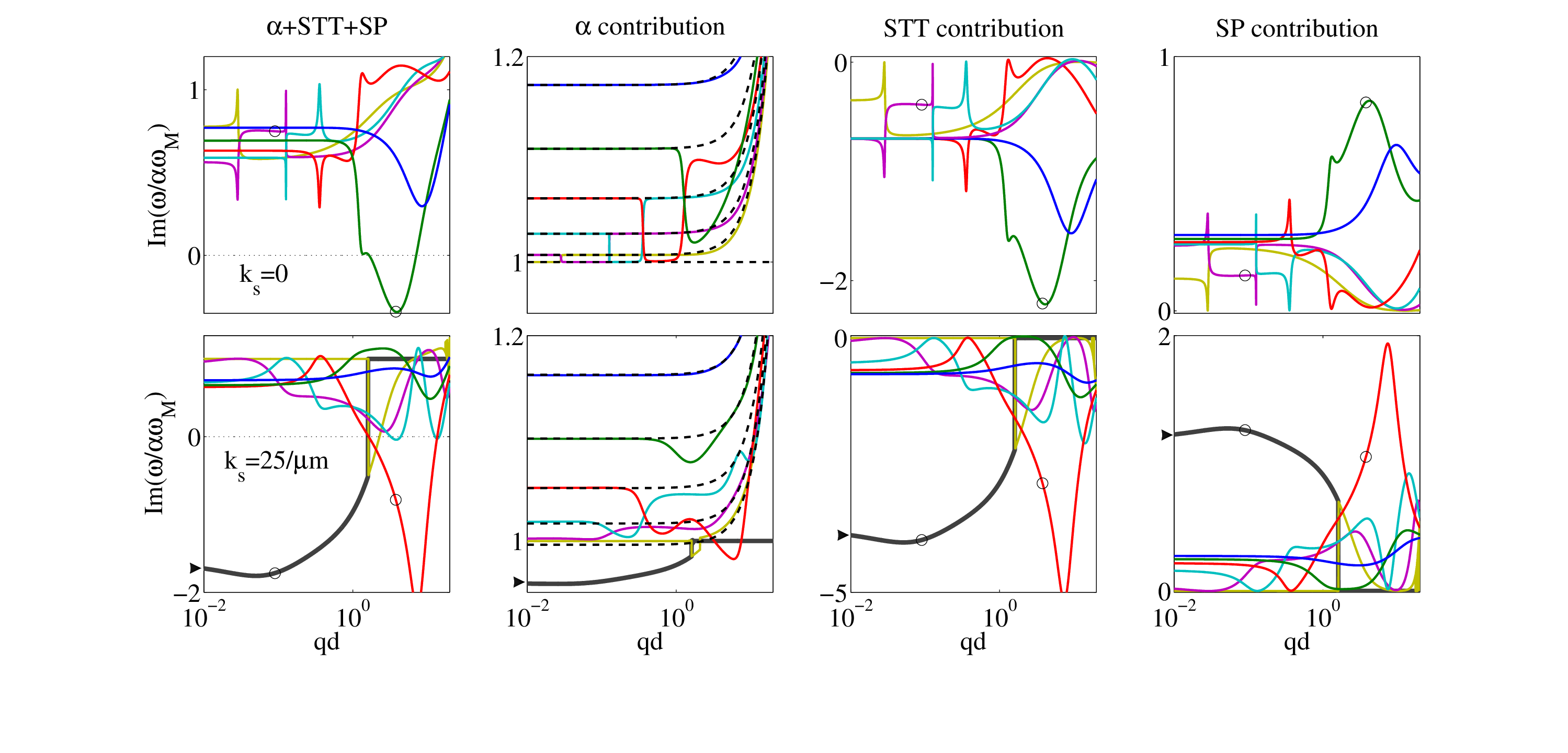}
    \caption{(Color online) Spin wave dissipation at $\theta = \pi/2$ (or $\qq\perp\mm$) with $g_r = 5.8\times 10^{17}$/m$^2$ ($k_p = 0.01/\upmu$m). Top row: no surface anisotropy ($k_s = 0$), bottom row: with easy-axis surface anisotropy ($k_s=25/{\upmu}$m). The 1st column is the total dissipation with current injection of $J_c = 2.3\times 10^{11}$A/m$^2$ ($k_j = 0.35k_c$). The 2nd column to 4th column are the contributions from the Gilbert damping, STT, and SP, respectively. For all panels, the solid lines (different colors denote different bands) are calculated from the numerical solution of \Eq{eqn:Mab}, and the dashed lines (and the $\blacktriangleright$) are plotted using the analytical expressions given by the imaginary parts of \Eqss{eqn:wn}{eqn:wS}.}
        \label{fig:SP_diss}
\end{figure*}
\subsection{Spin wave dispersion \& profiles}
\label{sec:swdispersion}

The spin wave dispersion, \ie the real part of the mode frequency $\re{\omega}(\qq)$, is plotted in \Figure{fig:SP_disper_ks0}(a) for $\theta = \pi/2$ (or $\qq\perp\mm$) when there is no surface anisotropy ($k_s = 0$). The dispersion can be separated into the dipolar spin wave regime for $qd\lesssim 1$, where the dispersion relation is flat (for $\theta =\pi/2$ only, non-flat for other angles), and the exchange spin wave regime for $qd > 1$, where the dispersion relation is approximately parabolic and increasing with $A_{\rm ex}q^2$. In the dipolar regime ($qd\lesssim 1$), there are multiple flat bands (associate with different transverse modes in the $x$ direction) and a magnetostatic surface wave (MSW) that crosses with the lowest four flat bands. These results are identical to our previous studies. \cite{de_wames_dipole-exchange_1970} The spin wave profiles for the typical dipolar/exchange spin waves ($qd = 0.09/3.78$) are shown in \Figure{fig:SP_disper_ks0}(b)/(c). For the dipolar spin waves (\Figure{fig:SP_disper_ks0}(b)), the bulk modes (corresponding to the flat bands) are simply the standing waves confined by the film thickness $d$. The MSW mode (thick purple curve in \Figure{fig:SP_disper_ks0}(b)) is a surface wave, but with a very long penetration depth, which means that the MSW mode for small $q$ is actually more like a uniform mode rather than a surface mode.

The more interesting physics happens when including the surface anisotropy $k_s$, which can take either sign: $k_s > 0$ means that the surface spins tend to align with the surface normal and is called easy-axis surface anisotropy (EASA), while $k_s < 0$ means that the surface spins tend to lie in the plane of the surface and is called hard-axis surface anisotropy (HASA). One effect of the surface anisotropy is to shift the bulk band frequencies as indicated by \Eq{eqn:wn}: the positive/negative $k_s$ shift the frequencies downwards/upwards. For EASA ($k_s>0$), as discussed in our previous study, \cite{xiao_spin-wave_2012} a new type of surface spin wave mode (the lowest thick black band in \Figure{fig:SP_disper_ks25}(a)) appears. The magnetization profile for this EASA surface wave at $qd = 0.09$ (the mode indicated by the circle on the thick black band in \Figure{fig:SP_disper_ks25}(a)) is plotted as the thick black curve in \Figure{fig:SP_disper_ks25}(b), which shows its surface feature. The penetration depth $d_s$ of the EASA surface wave is inversely proportional to the strength of the EASA: $d_s \propto 1/k_s$. \cite{xiao_spin-wave_2012} 

\subsection{Spin wave dissipation}
\label{sec:swdissipation}

The STT and SP mainly affect the dissipation of spin waves \ie the imaginary part of the mode frequency, and leave the spin wave dispersion and profiles discussed in the previous section practically unchanged. 

The spin wave dissipation, $\im{\omega}$, is plotted in the 1st column of \Figure{fig:SP_diss} for the two cases of surface anisotropy as those in \Figure{fig:SP_disper_ks0} and \Figure{fig:SP_disper_ks25}: $k_s = 0$ (top) and $k_s = 25/\upmu$m (bottom). In both plots, STT due to current injection $J_c = 2.3\times 10^{11}$A/m$^2$ and SP are included. The interfacial mixing conductance value is taken as $g_r = 5.8\times 10^{17}$/m$^2$. 

In linear response regime, different mechanisms for the spin wave dissipation are additive. As indicated by the analytical results \Eqss{eqn:wn}{eqn:wS} in Section \ref{sec:SW_analytical}, there are three different contributions to the dissipative imaginary part $\im{\omega}$: the Gilbert damping ($\alpha$ term), STT ($k_j$ term), and SP ($k_p$ term). We plot these contributions to $\im{\omega}$ separately in the 2nd-4th column in \Figure{fig:SP_diss}. The 2nd column, the Gilbert damping contribution, is equivalent to the dissipation for a  YIG film without Pt capping layer (thus no STT or SP). The 3rd and 4th columns are the contributions from STT and SP respectively, which show very similar $\qq$-dependence in shape but with opposite sign. Apart from an overall prefactor determined by the structure and material parameters ($\tau_{\rm stt}$ and $\tau_{\rm sp}$ in \Eq{eqn:torque}), the overall shape of STT and SP is determined by the interfacial transverse magnetization $\mm_\perp(0)$ (through the vectorial part of \Eq{eqn:torque}), which is strongly mode dependent (or $\qq$-dependent). This common ingredient for STT and SP leads to their similarities in the $\qq$-dependence. The sign is governed by the polarity of the charge current $J_c$.

When surface anisotropy is absent ($k_s = 0$, top panels in \Figure{fig:SP_diss}), the green band reaches negative dissipation for large $q$. This negativity is because the STT contribution reaches its (negative) maximum for the green mode at large $q$. Such large STT contribution is due to its large interfacial magnetization $\mm_\perp(0)$ for the green mode, which can be seen from its profile in the thick green curve in \Figure{fig:SP_disper_ks0}(c). On the opposite, the $\mm_\perp(0)$ for the red mode (\Figure{fig:SP_disper_ks0}(c)) is small, therefore the STT has little effect on the red mode at large $q$, this is why the STT contribution for the red mode is close to zero for $qd>1$. The SP contribution has the same feature as the STT because SP also depends on $\mm_\perp(0)$.

For the case with EASA ($k_s = 25/\upmu$m, bottom panels in \Figure{fig:SP_diss}), the features of large/small STT/SP contributions are due to the same reason as in the no surface anisotropy case that they all determined by the interfacial value $\mm_\perp(0)$ for a specific mode. The main difference between these two surface anisotropy cases is from the additional EASA surface wave (the lowest thick black band in \Figure{fig:SP_disper_ks25}(a)). Because of its strong localization near the interface, STT and SP strongly affect this mode, and the STT/SP contribution for this mode (the black curve in the bottom right two panels of \Figure{fig:SP_diss}) becomes  larger. For two typical modes indicated by circles on the black/red bands, the large STT and SP contributions are caused by their surface wave features, as observed in  their profiles (thick black/red curves in \Figure{fig:SP_disper_ks25}(b)/(c)).

Overall, STT and SP have a  larger effect on surface waves, such as the MSW (at larger $q$) and EASA surface waves. Therefore, in the absence of an applied current, the surface waves have larger damping due to larger SP contribution. When a large enough charge current is applied, the STT contribution overcomes that of the Gilbert damping and SP, and excites  preferably surface waves. 

\subsection{Power spectrum and threshold current}
\label{sec:power}

\begin{figure}[t]
\centering
                \includegraphics[width=\columnwidth,trim= 90 15 480 25, clip=true]{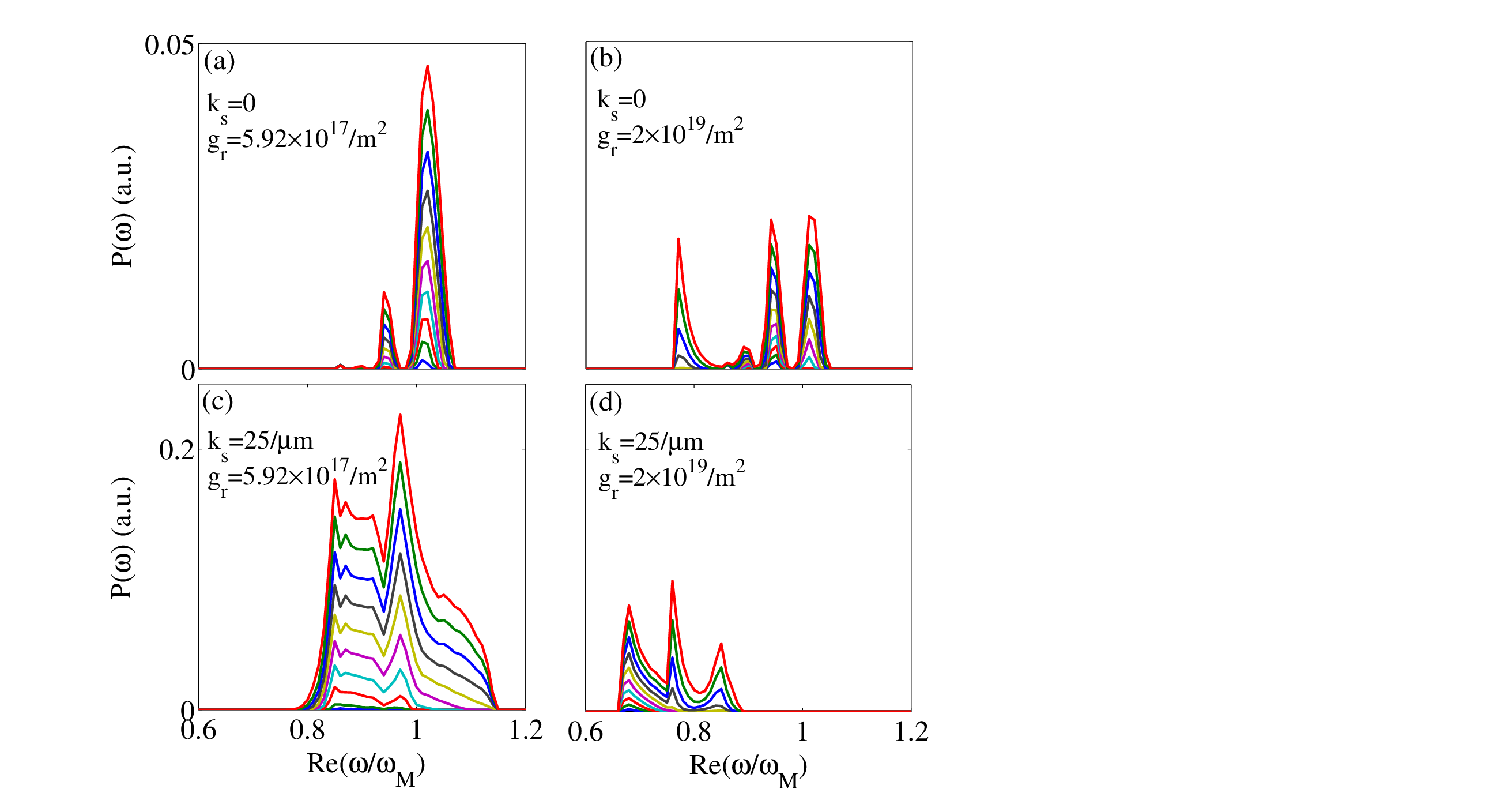}
        \caption{(Color online) Power spectrum (resolution ${\delta}{\omega}/{\omega}_M = 0.01$) for different combinations of surface anisotropy and mixing conductance at ten current levels (increasing by ${\delta}k_j = 0.01k_c$) above threshold current. }
        \label{fig:SP_spectrum}
\end{figure}

Since there are multiple spin wave modes excited simultaneously by the STT, we study the frequency dependence of the excitation power. Because the theory is based on linear response, we can only predict the onset of the excitation of a certain spin wave mode. Its tendency of being excited can be measured by the value of $\im{\omega}$: a more negative  $\im{\omega}$  implies more power. Therefore, we define an approximate power spectrum for the spin wave excitation:
\begin{equation}
        P({\omega}) = {\sum}_n{\int}_{\rm \im{{\omega}_{\it n}}<0} \abs{\rm \im{{\omega}_{\it n}(\qq)}} 
        {\delta}[{\omega}-\re{{\omega}_{\it n}(\qq)}]d\qq,
\label{eqn:Pw}
\end{equation}
which summarizes the information about the mode-dependent current-induced amplification as a sum over bands with band index $n$. \Figure{fig:SP_spectrum} shows the power spectrum computed from \Eq{eqn:Pw} for different surface anisotropies and mixing conductances. 


Let us first inspect the effect of EASA. As seen in \Figure{fig:Jc}(b) (the filled/empty dots are for with/without EASA), EASA reduces the threshold current by about a factor of two. In addition, EASA also greatly enhances the excitation power, as seen by the comparison between the top and bottom panels in \Figure{fig:SP_spectrum}. The reason for this effect is the strong confinement of the EASA mode (see thick black profile in \Figure{fig:SP_disper_ks25}(b)) and correspondingly low threshold current (given by \Eq{eqn:jc}). Almost all EASA modes in  $\qq$ phase space are  excited  simultaneously (see the lower panels of \Figure{fig:SP_diss}). Easy excitation and the large excitation phase space, lead to the large excitation power in the presence of  EASA. In comparison, for $k_s = 0$ the excitation threshold current is higher and the modes that can be excited  occupy only a small area of phase space (only a small window of the green band can be excited as seen in \Figure{fig:SP_diss}).     

It is also interesting to compare the power spectrum for different mixing conductances $g_r$. Comparing \Figure{fig:SP_spectrum}(a - b) for $k_s = 0$ (or \Figure{fig:SP_spectrum}(c - d) for $k_s = 25/\upmu$m), we observe that an increasing mixing conductance tends to shift the power spectrum to lower frequencies, or cause a red shift. Both the STT and SP depend on (or are proportional to) the mixing conductance $g_r$ (see \Eq{eqn:torque}) and the interfacial value of the transverse magnetization $\mm_\perp(0)$, which dominates the $\qq$-dependence. The SP also depends on the frequency $\dmm(0)$ and is more effective for the high frequency modes, while the STT does not depend explicitly on frequency. As a consequence, a large mixing conductance tends to suppress the excitation of high frequency modes, thereby causing a red shift of the power spectrum.

\section{Discussions \& conclusions} 
\label{sec:discussions}

The EASA induced surface wave mode for $k_s > 0$ has several properties which make this mode superior for spin information processing and transport: 1) it can be easily induced unintentionally or by engineering the surface anisotropy, 2) its penetration depth is controlled by the strength of the surface anisotropy, 3) it can be excited by relatively small currents, 4) it has a finite group velocity and can propagate long distances (in the absence of SP). The required surface anisotropy for this new surface mode is ubiquitous in magnets and sensitive to surface treatments and overlayers, which can be used advantageously, \textit{e.g.} to decorate the magnetic insulator surface to create corridors or circuits which can accommodate this surface wave mode and its propagation. 

We find a threshold current for spin wave excitation for Pt\textbar YIG structures to be in the range of $10^{10}\sim 10^{11}$A/m$^2$ for typical parameters (spin Hall angle $\theta_H = 0.08$, mixing conductance $g_r \simeq 10^{18}\sim 10^{19}$/m$^2$). This value is higher than the value predicted in Ref. \onlinecite{xiao_spin-wave_2012}, which assumes perfect spin current absorption at the interface and ignores the SP effect on the spin wave, while both tending to underestimate the threshold current. The theoretical value is much higher than the experimental value for the threshold current of $10^9$A/m$^2$, \cite{kajiwara_transmission_2010} (even when accounting for the EASA surface wave). Although there are uncertainties in the value of surface anisotropy, spin Hall angle, spin-flip length, \etc any/all of these cannot reconcile a discrepancy between the experiment and the theory of almost two orders of magnitude.

In summary, we presented a self-consistent theory for the current-induced magnetization dynamics in normal metals\textbar ferromagnetic insulators bilayer structure, including the effects of STT and SP at the interface. \cite{tserkovnyak_enhanced_2002} We found that 1) the mode dependence of the STT and SP scales identically and surface waves are more affected than bulk waves, 2) the SP causes a red shift in the power spectrum, and 3) easy-axis surface anisotropy can induce a new type of (EASA) surface wave mode, which typically has the lowest threshold current for excitation and contributes most to the excitation power. We propose that engineering the surface anisotropy and the EASA surface waves might facilitate applications in low power spintronic-magnonic hybrid circuits.
%
 
\section*{Acknowledgement}
We acknowledge support from the University Research Committee (Project No. 106053) of HKU, the University Grant Council (AoE/P-04/08) of the government of HKSAR, the National Natural Science Foundation of China (No. 11004036, No. 91121002), the Marie Curie ITN Spinicur, the Reimei program of the Japan Atomic Energy Agency, EU-ICT-7 "MACALO", the ICC-IMR, DFG Priority Programme 1538 "Spin-Caloric Transport", and Grand-in-Aid for Scientific Research A (Kakenhi) 25247056..

\bibliographystyle{apsrev}

\end{document}